\documentclass[useAMS,usenatbib,twocolumn]{mn2e}
\usepackage{amsfonts}
\usepackage{amsmath}
\usepackage[dvips]{graphicx}
\usepackage{epsfig}
\usepackage{wrapfig}
\usepackage{amssymb}
\usepackage{longtable}
\usepackage{accents}
\usepackage{dcolumn}
\usepackage{bm}
\usepackage[caption=false]{subfig}

\begin{document}

\title[SMBH binary in PG~1302--102]{Constraining the parameters of the putative supermassive binary black hole in PG~1302--102 from its radio structure}
\author[E. Kun, S. Frey, K. \'{E}. Gab\'{a}nyi, S. Britzen, D. Cseh, L. \'{A}. Gergely]{E. Kun$^{1,2}$\thanks{%
E-mail: kun@titan.physx.u-szeged.hu}, S. Frey$^{3}$, K. \'{E}. Gab\'{a}nyi$^{3,4}$, S. Britzen$^{5}$, D. Cseh$^{6}$, L. \'{A}. Gergely$^{1,2}$ \\
$^{1}$Department of Experimental Physics, University of Szeged, D\'om t\'er 9, H-6720 Szeged, Hungary\\
$^{2}$Department of Theoretical Physics, University of Szeged, Tisza Lajos krt 84-86, H-6720 Szeged, Hungary\\
$^{3}$F\"{O}MI Satellite Geodetic Observatory, P.O. Box 585, 1592 Budapest, Hungary\\
$^{4}$Konkoly Observatory, MTA Research Centre for Astronomy and Earth Sciences, P.O. Box 67, 1525 Budapest, Hungary\\
$^{5}$Max-Planck-Institute f\"{u}r Radioastronomie, Auf dem H\"{u}gel 69, D-53121 Bonn, Germany\\
$^{6}$Department of Astrophysics/IMAPP, Radboud University Nijmegen, PO Box 9010, NL-6500 GL Nijmegen, the Netherlands}
\date{Accepted . Received ; in original form }
\maketitle

\begin{abstract}
We investigate the pc-scale kinematics and kpc-scale radio morphology of the quasar PG~1302--102, which may harbour a sub-pc separation supermassive binary black hole system at its centre as inferred from optical variability. High-resolution radio interferometric measurements obtained with the Very Long Baseline Array (VLBA) in the Monitoring Of Jets in Active galactic nuclei with VLBA Experiments (MOJAVE) programme at 15~GHz at 20 epochs spanning 17 years were analysed to investigate the pc-scale radio structure. Archival observations with the Very Large Array (VLA) at $1.4$~GHz and $5$~GHz were obtained to study the kpc-scale morphology. We find that the pc-scale jet is inclined within $\sim 2.2\degr$ to the line of sight and has a half-opening angle of $\sim 0.2\degr$. The parameters derived from the pc-scale radio jet are qualitatively consistent with those obtained from the analysis of the optical light curve of PG~1302--102. We obtain at least $0.08$ for the mass ratio of the two black holes in the system. We find some indication for a helical jet structure on kpc-scale, but the directions of the inner and the extended radio jets are significantly different, obstructing a straightforward connection of the pc- and kpc-scale jets within the binary scenario.
\end{abstract}

\pagerange{\pageref{firstpage}--\pageref{lastpage}}

\label{firstpage}


\section{Introduction}

The quasar PG~1302--102 came into focus since \citet{Graham2015} presented its optical variability with a period of $5.2\pm0.2$~yr. This was found in a systematic search for variable sources in the $V$-band light curves of the Catalina Real-Time Transient Survey (CRTS). They interpreted this period as a consequence of the orbital motion of a supermassive binary black hole (SMBBH), and suggested that PG~1302--102 is a tight SMBBH candidate with a separation of about 0.01~pc. PG~1302--102 is a bright, flat-spectrum radio quasar at redshift $z=0.278$ \citep{Marziani1996} hosted by an elliptical galaxy, as typical for radio-loud quasars.

The existence of SMBBHs is a natural consequence of the hierarchical galaxy evolution where the galaxies and their central black holes (BH) grow via major mergers \citep[e.g.][]{Kauffmann2000}. In the first phase of the evolution, field stars carry away the energy of the binary black hole (BBH) and the binary shrinks under the control of the dynamical friction. After the depletion of stars in the nuclear region, an intermediate phase follows. The binary spends the majority of its lifetime at separations of $0.01\mbox{--}1$~pc \citep{Begelman1980}. In the third phase, the gravitational radiation takes over in the energy and orbital angular momentum dissipation, gradually leading to a merger. The sub-pc separation of the putative SMBBH in PG~1302--102 \citep{Graham2015} suggests that this binary already evolved into the third, inspiral phase of the merger. For a recent review on the compact SMBBHs from observational point of view see \citet{Komossa2015}.

Gravitational waves emitted by sub-pc separation SMBBHs would be observable with Pulsar Timing Arrays \citep[PTA; e.g.][]{Sesana2009,Manchester2013}, with sensitivity on the low-frequency regime of gravitational waves. Identification of possible targets is essential to constrain the parameter field of the gravitational waveform in order the detect them. Based on the current parameter estimates, PG~1302--102 unfortunately remains by a factor of $\sim10$ below the detection threshold of even the future PTAs \citep{Orazio2015}.

PG~1302--102 belongs to the subclass of radio-loud quasars representing only about ten per cent of the optically identified quasars \citep[e.g.][]{Kellermann1989,Stocke1992}. Studies suggest that a fast rotating BH might give rise to relativistic jets emanating from the core \citep{Penrose1969,BlandfordZnajek1977}, causing the radio loudness of the source via synchrotron radiation of the charged particles spiralling in the magnetic field with relativistic speeds. It is plausible to suspect that BHs in radio-loud quasars have high spin. Probably the SMBH spin is the key factor to explain the dichotomy between radio-quiet and radio-loud quasars \citep[e.g.][]{Tchekhovskoy2010}.

The detailed analysis of the radio jet morphology in active galactic nuclei (AGN) has the potential to unveil the properties of the central engine. The technique of Very Long Baseline Interferometry (VLBI) is successfully applied to investigate the inner structure of the jets with the finest resolution. The Monitoring Of Jets in Active galactic nuclei with VLBA Experiments{\footnote{\tt http://www.physics.purdue.edu/astro/MOJAVE/}} (MOJAVE) offers calibrated VLBI visibilty data of jets associated with AGN at the observing frequency of $15$~GHz. PG~1302--102 was also monitored and calibrated, and its data are available from the MOJAVE Survey \citep{Lister2009}.

 Aiming to check whether the Very Long Baseline Array (VLBA) observations reveal fingerprints of the BBH on the jet structure and motion on milliarcsecond (mas) angular scale, we decompose the brightness distribution of the jet into Gaussian components at 20 different epochs over a time interval of more than 17 years, and investigate their long-term motion. We also use two archival Very Large Array (VLA) data sets of PG~1302--102 at 1.4 and 5~GHz observing frequencies, to investigate the jet morphology at arcsec-scale. A slow precession of the spin of the BH responsible for launching the jet may manifest itself in a helical structure of the large-scale radio jet.

In this paper, we investigate and interpret the radio structure of PG~1302--102 both on pc- and kpc-scales. The paper is organized as follows. In Section \ref{datareduction}, we describe the brightness distribution model fits to archival VLBA and VLA observations. In Section \ref{morphology}, we discuss the morphology and kinematics of the jet of PG~1302--102. In Section \ref{smbbh}, we interpret our results and constrain the parameters of the SMBBH. In Section \ref{remarks}, we give concluding remarks and a summary. A flat $\Lambda\mathrm{CDM}$ cosmological model is adopted throughout the paper, with Hubble constant $H_0=67.8\mbox{~}\mathrm{km\mbox{~}s^{-1}\mbox{~}Mpc^{-1}}$ and matter density $\Omega_\mathrm{m}=0.308$ \citep{Planck2015}. In this model, the angular scale is $4.352$~kpc~arcsec$^{-1}$ at the redshift of the source.

\begin{figure*}
    \includegraphics[scale=0.8]{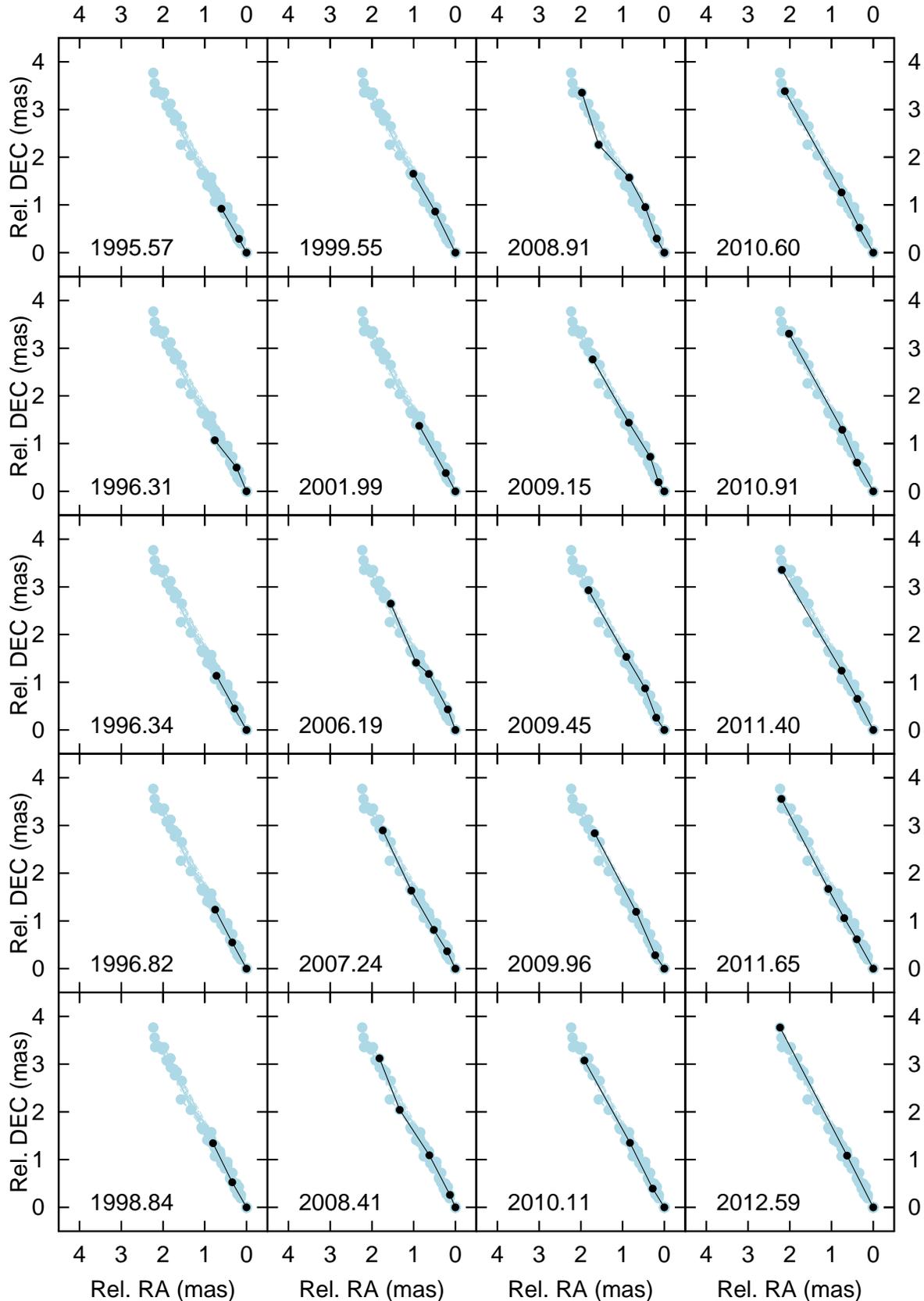}
    \caption{Temporal variation of the PG~1302--102 jet ridge line across the observing epochs of the MOJAVE radio observations with the VLBA at $15$~GHz. Relative right ascension and relative declination are shown on the horizontal and vertical axes, respectively. The VLBI core lies at the ($0$,$0$) position. The black dots denote the positions of the peak of the Gaussian components at a given date indicated in the bottom-left corner of each panel. The light dots represent all components at all epochs in each of the panels.}
    \label{fig:allxy}
\end{figure*}

\begin{figure*}
    \includegraphics[scale=0.85,angle=270]{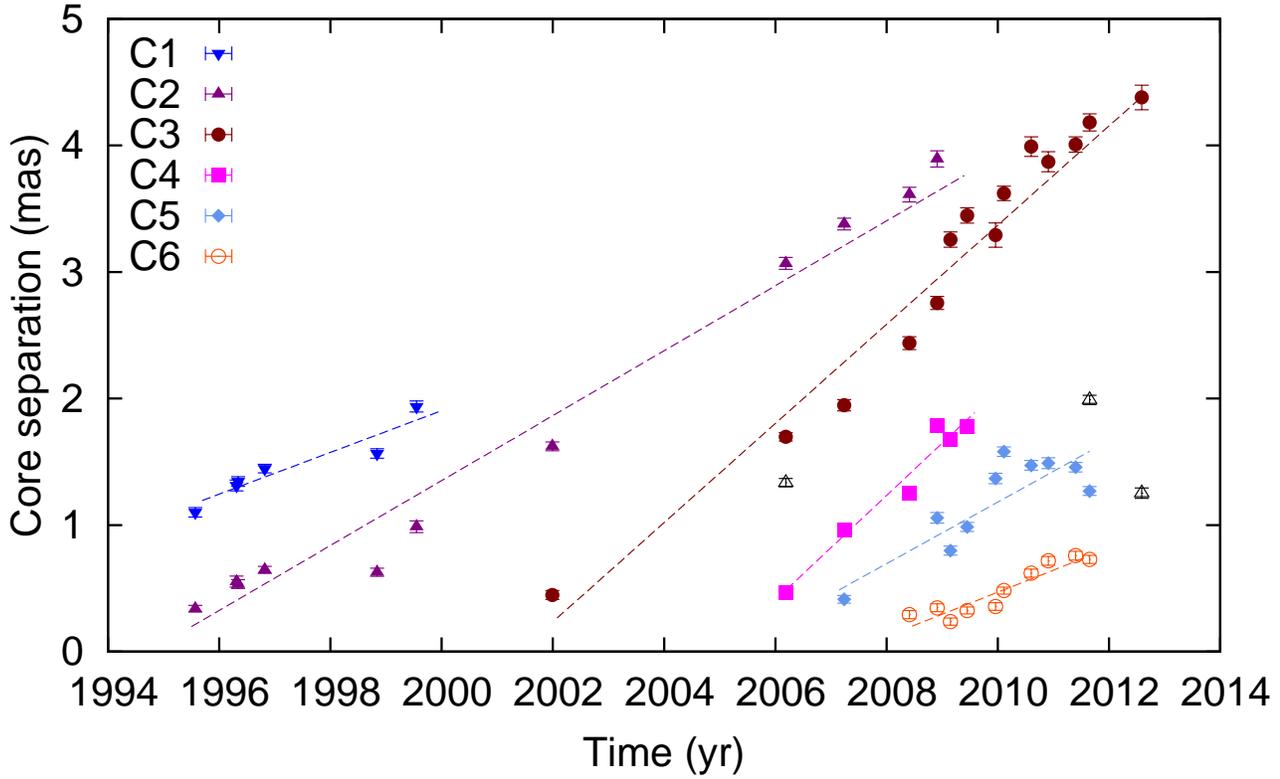}
    \caption{Component separations from the core as a function of time. The components identified through different epochs (C1, C2, C3, C4, C5, C6) are shown with different symbols. The dashed lines indicate their fitted linear proper motions. A few unidentified components not used for the proper motion determinations are marked with open up-pointing triangles.}
    \label{fig:coresep}
\end{figure*}

\begin{figure*}
    \includegraphics[scale=0.7,angle=270]{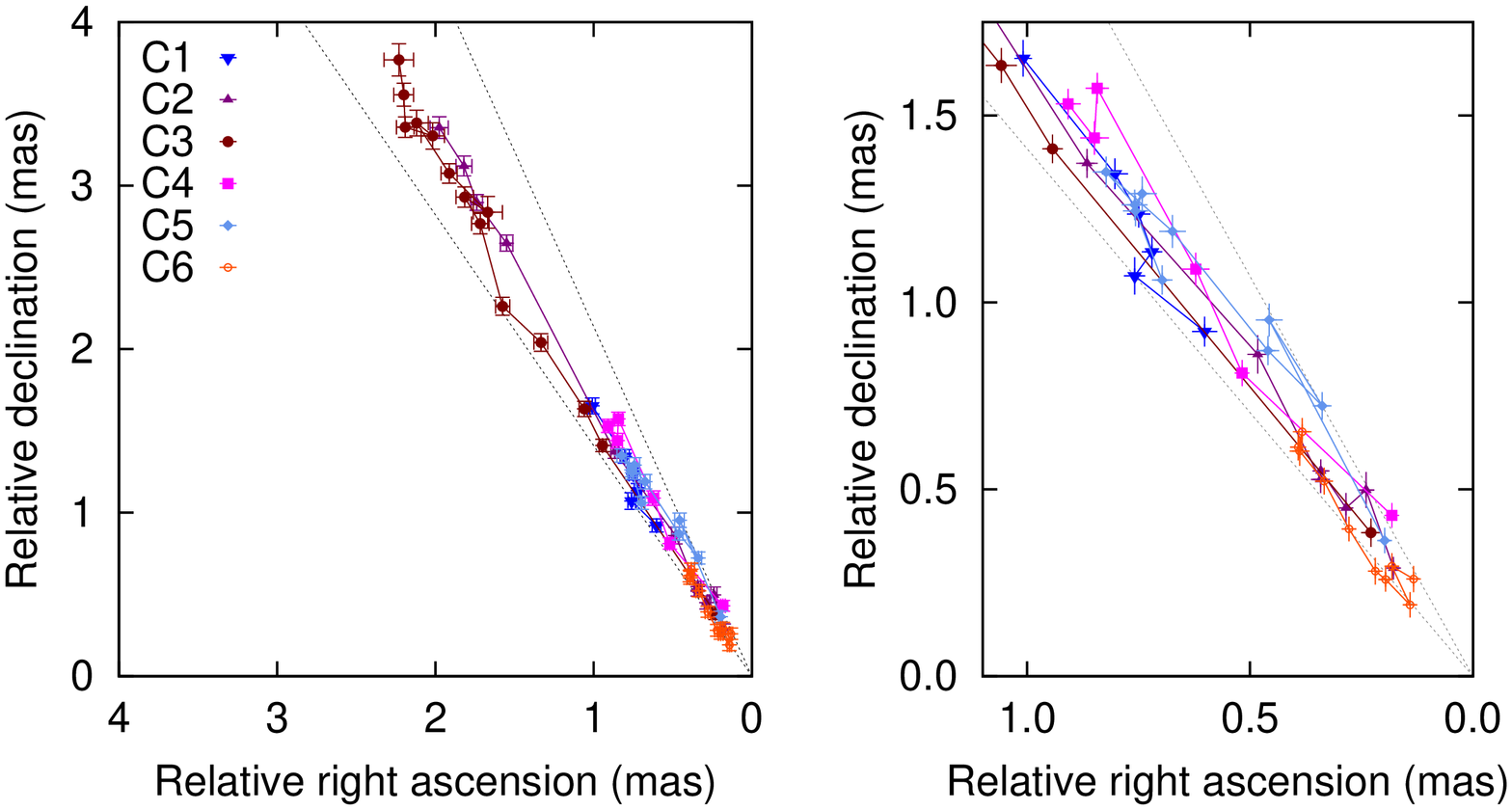}
    \caption{Relative right ascension and relative declination of the components from C1 to C6 (see in Fig.~\ref{fig:coresep}), at all epochs. The exploded view of the innermost part is in the right panel. The estimated jet opening angle is marked with the dotted lines.
}
    \label{fig:xy_together}
\end{figure*}

\section{Radio interferometric data}
\label{datareduction}

\subsection{Gaussian decomposition of the archival VLBA data}

Calibrated VLBI visibilities of PG~1302--102 are available from the MOJAVE database \citep{Lister2009}, observed in the period from $1995.57$ till $2012.59$ at $20$ epochs. We used these data to reveal the radio structure at pc-scale. The standard modelfit procedure in the Caltech {\sc Difmap} program \citep{Shepherd94} was used to decompose the brightness distribution of the jet into circular Gaussian components at each individual epoch. The fitted parameters of the circular Gaussians are the integral flux density, the position of the component, and its diameter (full width at half maximum, FWHM). The error estimates for the fitted parameters were obtained in the same way as described in details in \citet{Kun2014}. Component identification was performed using the modelfit results, in a way that only slight shifts were allowed in the component parameters between subsequent epochs. The brightest component at all epochs but 1999.55, when a newly emerged component became temporarily brighter, was identified with the ``core'' and was considered as the reference point. The right ascension and declination of the jet components are given relative to the position of the VLBI ``core'' (Fig.~\ref{fig:allxy}). The component separations from the ``core'' as a function of time are plotted in Fig.~\ref{fig:coresep}, their relative right ascensions and declinations at all epochs together are shown in Fig.~\ref{fig:xy_together}. 

\subsection{Calibration and Gaussian decomposition of the archival VLA data}

Deep 1.4-GHz VLA observations of PG~1302--102 were obtained from the US National Radio Astronomy Observatory (NRAO) archive\footnote{\tt http://archive.nrao.edu}. The data were taken in the most extended A configuration of the array, with a bandwidth of 100~MHz, on 1992 October 9-10 (project code: BC012). The total on-source integration time was nearly 1~h. Phases and amplitudes were calibrated in the Astronomical Image Processing System \citep[{\sc AIPS},][]{Greisen03} in a standard way, except that none of the primary VLA flux density calibrators were scheduled in the experiment. Therefore the source OQ~208 (J1407+2827) was used to define the flux density-scale, by adopting the total flux density value of $817$~mJy obtained from the NRAO VLA Sky Survey \citep[NVSS,][]{Condon98}. OQ~208 is known to be a compact but somewhat variable source \citep[e.g.][]{Wu13}. Because of the uncertainty in the absolute flux density calibration, we assume that the image brightness scale can be offset by up to $\sim$15 per cent. However, this does not affect our results, since we use the component positions in the brightness distribution model but not the flux densities in our subsequent analysis. The 1.4-GHz VLA image of PG~1302--102 shown in Fig.~\ref{fig:vla_helixfit} was made in {\sc Difmap} \citep{Shepherd94} from the calibrated visibility data exported from AIPS, using a standard hybrid mapping process.

We also obtained archival 5-GHz VLA A-configuration data taken on 1992 November 18 (project code: AG361). The total bandwidth was 100~MHz. The source PG~1302--102 was observed for 35 min. Phases and amplitudes were again calibrated in {\sc AIPS}. The primary VLA flux density calibrator 3C\,286 was used to define the amplitude scale. Imaging and model-fitting were performed in {\sc Difmap}. The 5-GHz image (not reproduced here) shows qualitatively the same two-sided arcsec-scale jet structure as the 1.4-GHz one, also known from the literature \citep[e.g.][]{Hutchings1994}. The extended jet components are less pronounced here than at 1.4~GHz, due to the higher angular resolution and the steep spectrum.

\section{Radio structure of PG~1302--102}
\label{morphology}

While VLBI is able to map compact features and their variations in radio sources, its typical resolution is still far from being sufficient to directly image such a tightly bound SMBBH that is expected in the centre of PG~1302--102 \citep{Graham2015}. Therefore we focus on an indirect way to identify the presence of SMBBH, and analyse long-term radio data to investigate the structure and kinematics of the jet.

\subsection{PG~1302--102 on pc scale}

Long-term VLBI monitoring data offer a way to check the validity of the scenario of a sub-pc separation SMBBH as being the central engine of PG~1302--102. Parameters of Gaussian components fitted to archival VLBA data from the MOJAVE survey span $17.02$ years in $20$ epochs. This period is long enough, and the time sampling is sufficiently dense to allow us to investigate the jet kinematics. 

To see whether the jet ridge line is helically disturbed, we plot the relative positions of the Gaussian components in Fig.~\ref{fig:allxy}. If we look at the jet as a whole, its ridge line is rather straight and not curved. The jet points nearly to the same direction across the MOJAVE epochs. The average position angle of the jet is $\Theta_\mathrm{pc}=31\fdg6\pm0\fdg6$ (position angles are conventionally measured from north through east). 

We identified $6$ components of the brightness distribution of the jet and followed their motions through multiple epochs. Their core separations versus time are plotted in Fig.~\ref{fig:coresep}. 
All components show apparent superluminal motion, suggesting small inclination angle and large bulk speed of the pc-scale jet. In the linear approximation, the components show different proper motions. The maximum proper motion (i.e. the steepest slope among the lines in Fig.~\ref{fig:coresep}) belongs to the component C4, $\mu^\mathrm{max}=0.41\pm0.04$~mas~yr$^{-1}$. It is translated to the maximum apparent speed of $\beta_\mathrm{app}^\mathrm{max}=7.5\pm0.7$, expressed in the units of the speed of light $c$. Then the minimum bulk Lorentz factor is constrained as $\Gamma^\mathrm{min}=6.8$, and the minimum jet velocity as $\beta_\mathrm{j}^\mathrm{min}=0.989$.

We plot the relative right ascension and declination of the components collectively in Fig.~\ref{fig:xy_together}. Interestingly, the components seem to be launched in slightly different directions, but after about the first 1~mas they become collimated with each other. The collimation is the most prominent for the two components at the opposite edges of the jet, C1 and C5. 
We use the outermost positions of C1 and C5 components (within $\sim$1.5~mas separation from the core) to define the opening angle of the jet cone as shown by the dotted lines in Fig.~\ref{fig:xy_together}. The apparent half-opening angle is $\psi_\mathrm{app}=5\fdg15\pm0\fdg75$.

A consequence of the Doppler boosting in a jet being inclined close to our line of sight is that one measures enhanced apparent brightness temperature $T_\mathrm{b}$ instead of the intrinsic brightness temperature $T_\mathrm{int}$. In case of VLBI components, the brightness temperature is calculated as \citep[e.g.][]{Condon1982}:
\begin{flalign}
T_\mathrm{b}=1.22\times 10^{12} (1+z) \frac{S}{\theta^2 f^2}\mbox{~}\mathrm{K},
\end{flalign}
where $S$~(Jy) is the flux density, $\theta$~(mas) is the FWHM diameter of the circular Gaussian component, and $f$~(GHz) is the observing frequency. The apparent and intrinsic brightness temperatures are related to each other as $T_\mathrm{b}=\delta T_\mathrm{int}$, where
\begin{equation}
\delta=\frac{1}{\Gamma(1-\beta_\mathrm{j} \cos \iota)}
\label{eq:Dopplerfactor}
\end{equation}
is the Doppler factor, $\beta_\mathrm{j}$ is the jet speed in the units of $c$ on pc scale, $\Gamma=(1-\beta_\mathrm{j}^2)^{-0.5}$ is the Lorentz factor, and $\iota$ is the inclination angle with respect to the line of sight. 

We calculated the apparent core brightness temperatures from the fitted Gaussian component flux densities and diameters at each of the 20 epochs of MOJAVE data. Assuming the equipartition brightness temperature $T_\mathrm{eq} \approx 5\times10^{10}$~K as the intrinsic brightness temperature \citep{Readhead1994}, we obtained the Doppler factors. For the following, we adopt the median value, $\delta=18.5\pm3.6$, as the Doppler factor characteristic to the source.

Then $\iota$ and $\beta_\mathrm{j}$ can be expressed from Eq.~\ref{eq:Dopplerfactor}, and also from the definition of the apparent speed
\begin{flalign}
\beta_\mathrm{app}\!=\!\frac{\beta_\mathrm{j} \sin\iota}{1-\beta_\mathrm{j} \cos \iota}.
\label{betaapp}
\end{flalign}
Substituting the measured Doppler factor $\delta$ and the proper motion in the jet $\beta_\mathrm{app}^\mathrm{max}$, we obtain $\beta_j=0.996^{+0.001}_{-0.002}$, $\Gamma=10.8^{+1.7}_{-1.9}$, and $\iota=2\fdg2\pm0\fdg5$. Then the jet inclination angle and the apparent half-opening angle determine the intrinsic half-opening angle of the jet cone as $\zeta=\psi_\mathrm{app}\sin\iota=0\fdg20\pm0\fdg05$.

The compact pc-scale radio structure of PG~1302--102 revealed by VLBI observations is a result of relativistic beaming and therefore the jet inclination is small. \citet{Graham2015} fitted the optical light curve of the PG~1302--102 assuming that its variability is caused by the periodically varying Doppler boosting in the jet. They obtained the following best-fitting parameters: Lorentz factor $\Gamma_\mathrm{opt}=5.4\pm0.1$, intrinsic half-opening angle $\zeta_\mathrm{opt}=0\fdg5\pm0\fdg1$, inclination angle $\iota_\mathrm{opt}=5\fdg0\pm0\fdg2$. Considering that \citet{Graham2015} used optical data with an assumed value of the spectral index, and our results are estimated independently from high-resolution MOJAVE radio data adopting a commonly used value for the intrinsic temperature of the VLBI core, the two sets of parameters are qualitatively consistent with each other. We can thus conclude that the pc-scale radio structure is compatible with the SMBBH model proposed by \citet{Graham2015}.

The analysis of the pc-scale morphology reveals that the jet length in PG~1302--102 is increasing, as seen from the time series plot of the jet components in Fig.~\ref{fig:allxy}. This suggests continuously strengthening jet activity. At the last MOJAVE epoch (2012.59), the total flux density of the jet components was $1.453$~Jy, almost three times higher than the average value before this epoch, indicating a radio outburst.
 
\begin{figure}
    \includegraphics[scale=0.7,bb=0 50 300 530]{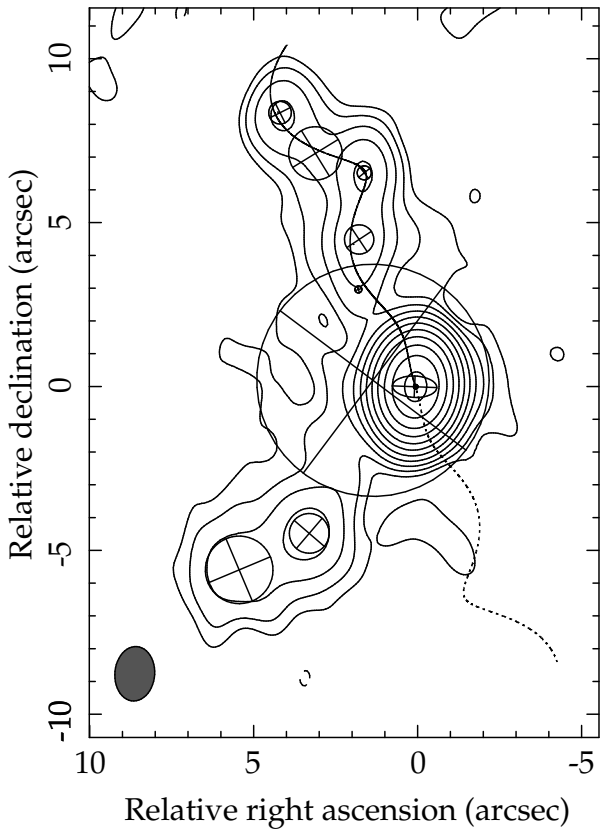}
    \caption{The 1.4-GHz VLA image of PG~1302--102, overplotted with the conical helix fitted to the components of the northern jet only. The dashed curve to the south of the core indicates the trajectory of a hypothetical jet which would be symmetric to the helical model fitted to the northern part. The real structure of the southern jet is markedly different from what would be expected from this simple model. The circles and the central ellipse indicate the Gaussian model components fitted to the visibility data in {\sc Difmap}. Their sizes represent the FWHM diameter. The image peak brightness is $543$~mJy~beam$^{-1}$. The lowest contour level is $\pm4.3$~mJy~beam$^{-1}$, further positive contours increase by a factor of 2. The off-source rms noise is $0.08$~mJy~beam$^{-1}$. The restoring beam size ($1\farcs66 \times 1\farcs21$) is shown in the bottom-left corner.}
    \label{fig:vla_helixfit}
\end{figure}

\subsection{PG~1302--102 on kpc scale}

The VLA image and the fitted Gaussian brightness distribution model components of PG~1302--102 (Fig.~\ref{fig:vla_helixfit}) are reminiscent of a helical structure in the northern jet, and a bent trajectory in the southern jet. At 1.4~GHz, the ``core'' and the northern jet is represented by 7 components, while the southern jet is described with just two components. The latter do not allow us to fit any meaningful geometric model. On the other hand, by assuming that the positions of the Gaussian components in the northern jet follow a helical geometry, we could fit a simple precession model to the data \citep[see rotation matrices in][]{Kun2014}. For this fit, the positions of the northern jet components from the 5-GHz VLA data were also used. This model provides estimates for the orientation and the half-opening angle of a helical jet.

The best-fitting parameters were obtained by iteratively performing non-linear least squares estimates using the Levenberg--Marquardt algorithm, such that the $\chi^2$ was minimized during the process. This way we got the position angle of the jet axis as $\Theta_\mathrm{kpc}=20\fdg15\pm0\fdg26$, the inclination of the jet axis with respect to the line of sight as $\iota_\mathrm{kpc}=65\fdg5\pm2\fdg2$, and the intrinsic half-opening angle of the jet as $\Psi^\mathrm{int}=6\fdg5\pm1\fdg0$. The axial growth of the jet during one spatial period is $a=6\farcs0\pm0\farcs1$, and the radial growth during one spatial period is $b=0\farcs7\pm0\farcs1$ (reduced $\chi^2=4.6$). Our results are qualitatively consistent with the two-sided appearance of the kpc-scale jet which also suggests large inclination angle.

We show the best-fitting northern helical jet plotted together with the 1.4-GHz VLA total intensity image and the component locations in Fig.~\ref{fig:vla_helixfit}. It is important to realize that the components themselves are not supposed to move along this helix, rather their instantaneous positions form a helix that is expanding as the components move away from the core on their ballistic paths. The dashed curve in Fig.~\ref{fig:vla_helixfit} indicates how a symmetrical helical counterjet would appear in the radio image of PG~1302--102 to the south of the core. The structure of the southern jet is clearly different, indicating that this simple precessing jet model cannot be applied to describe it. The asymmetric kpc-scale extended radio structure in this object could perhaps be explained by jet bending due to jet--cloud interactions in the host galaxy. 

\section{Interpretation}
\label{smbbh}

\subsection{Possible precession time scale of the kpc jet}

We estimate the precession period of the northern kpc-scale jet, assuming that the components indeed trace out a helical structure (Fig.~\ref{fig:vla_helixfit}). The physical expansion rate of the helical jet pattern is described by the deprojected axial ($a$) and radial growth ($b$) of the jet during one spatial period. Then the precession period in the rest frame of the source is
\begin{flalign}
T_\mathrm{j}=\frac{\sqrt{a^2+b^2}}{c \, \beta_\mathrm{j,kpc}}\frac{d_\theta}{1+z}=\frac{(6.74\pm1.05)\times 10^4}{\beta_\mathrm{j,kpc}}\mbox{~}\mathrm{yr},
\end{flalign}
where $\beta_\mathrm{j,kpc}$ is the jet speed on kpc scale, and $d_\theta=4.352$~kpc~arcsec$^{-1}$ is the angular scale at the redshift of the source. An order-of-magnitude estimate of the precession period is possible if we assume $\beta_\mathrm{j,kpc} \approx 0.1$ for the kpc-scale jet expansion \citep[see e.g.][]{Scheuer1995}. In this case, $T_\mathrm{j}$ is nearly $7 \times 10^5$~yr. 

\subsection{The connection between the pc- and kpc-scale radio structures}

Contrary to the pc-scale jet, the existence of the extended two-sided kpc-scale radio structure in the core-dominated source PG~1302--102 indicates large jet inclination. However, the structure of the (northern) kpc-scale jet characterised by our helical model (Fig.~\ref{fig:vla_helixfit}) cannot be straightforwardly explained in the context of the same binary BH system. The intrinsic half-opening angle of the kpc-scale jet ($\Psi^\mathrm{int}\approx6\fdg5$) is much smaller than the difference between the inclination of the kpc-sclale ($\iota_\mathrm{kpc}\approx65\fdg5$) and pc-scale ($\iota\approx2\fdg2$) jets. This rules out the scenario where the pc-scale jet slowly precesses about the symmetry axis of the kpc-scale jet, within its cone. In other words, assuming that the pc-scale jet points to the direction of the spin of the BH responsible for the jet launching, the kpc-scale structure cannot be interpreted simply with the precession of the black hole spin in the binary system described by \citet{Graham2015}. 

However, the kpc-scale helical structure, if real, could possibly be reconciled with the SMBBH model if a significant jet reorientation, i.e. a spin-flip \citep{Gergely2009} happened in a transition phase between the production of the kpc- and pc-scale jets. We will further elaborate this possibility in Section~\ref{smbbh-3} and find it unlikely based on the derived BBH parameters of the system. An alternative, albeit more speculative explanation is that the outer and inner structures were produced by different black holes of the binary system. They might have changed their activity status in a way that the one responsible for the kpc-scale jet ``switched off'', and the other one now producing the pc-scale jet ``turned on'' the jet production fuelled by matter accretion. In this scenario, the different jet orientations are naturally explained by the different spin directions of the two central engines.

Another, similarly speculative scenario involves the interaction of three BHs, where initially two BHs merged and reached the final coalescence. In this picture, the kpc-jet would show the imprint of the interaction of the two BHs, revealing the spin-orbit precession of the dominant spin. Then a spin-flip reorients the spin, causing a sudden directional change of the jet, and the final coalescence happens. The newly-merged BH catches a third BH and emanates the jet seen in pc scale. This jet unveils the orbital motion of the newly-formed binary. However, considering the estimated precession period of the kpc-scale jet as the consequence of spin-orbit precession, the gravitational lifetime would be several million years \citep[e.g.][]{Gergely2009}, ruling out the possibility of a recent final merger in the system.

In an alternative three-body interaction scenario, the third BH could have been kicked out from the system due to a gravitational slingshot effect, rather than being absorbed. This is similar to one of the proposed explanations for the object CID-42 \citep{Civoni2010}. In this case, the orbital angular momentum could have been transferred to the current primary BH, causing a spin flip without invoking a recent BH merger within the past several million years. Simulations beyond the scope of this paper would be needed to prove the feasibility of this scenario.

We may also try to explain the kpc-scale helical structure assuming that it results from the precession of the accretion disk of the primary BH, induced by the disk-crossing motion of the secondary BH \citep{Caproni2004}. However, it would imply that the distance between the outer edge of the accretion disk and the primary BH is $100\mbox{--}1000$ times smaller than the orbital separation, ruling out the physical interaction between the disk and the secondary BH.

Finally, it is also possible that the northern jet is in fact not helical, and therefore is not shaped by precession. Even in this case, the apparent discrepancy between the relatively large jet inclination on kpc scale on the one hand, and on pc scale on the other hand needs to be explained. Asymmetries in the dense interstellar medium surrounding the core, and jet--cloud interactions may result in the deflection of the jet flow \citep[e.g.][]{Stocke1985} in a different way on both sides of the core. Such interactions were suspected by \citet{Hutchings1994} who analysed the spatial correlation of the high-resolution optical and radio structures in PG~1302--102.

\subsection{Binary parameters}
\label{smbbh-3}
We use the BBH-in-jet model presented in our earlier work \citep{Kun2014} to derive parameters of the putative SMBBH in PG~1302--102. The underlying process revealing the presence of the binary is that the orbital motion of the jet-emitter BH modifies the average inclination angle of the jet launching on the orbital time scale. Using time-travel argument, this affects the appearance of the pc-scale jet. If the binary has already evolved into the inspiral phase, a slow spin-orbit precession may be present as well. Its period is much longer than the orbital period, and its signal could be revealed typically on the kpc-scale jet. 

\citet{Graham2015} estimated the total mass of the central object in PG~1302--102 from various emission lines. Averaging these estimates, $m\approx4\times 10^8\mbox{~}m_\odot$. Considering it as the total mass of the binary, and the rest-frame optical period $T=4.0\pm0.2$~yr \citep{Graham2015} as the orbital period, the orbital separation is $r\approx0.01$~pc. Then the post-Newtonian parameter $\varepsilon=Gm r^{-1} c^{-2}\approx0.002$ suggests that the binary progressed into the inspiral phase of the merger ($G$ denotes the gravitational constant).

We can estimate the mass ratio of the two BHs, $\nu=m_2/m_1$ ($m = m_1 + m_2$; $m_1>m_2$), using the orbital velocity of the dominant BH that we assume as the jet emitter:
\begin{flalign}
v_\mathrm{orb} = \left( \frac{Gm}{r} \right) ^{\frac{1}{2}} \frac{\nu}{1+\nu}.
\label{vorb} 
\end{flalign}
The intrinsic half-opening angle of the cone along which the inclination angle of the jet changes is expressed as \citep{Kun2014}:
\begin{flalign}
\zeta = \arcsin \frac{v_\mathrm{orb}\cos \kappa}{c \, \beta_\mathrm{j}}, 
\end{flalign}
where $\kappa$ is the angle between the Newtonian angular momentum and the dominant spin. Substituting the expression of the orbital velocity from Eq.~\ref{vorb},
\begin{flalign}
\kappa = \arccos \frac{\beta_{\mathrm{j}} \sin \zeta}{\varepsilon^{1/2}} \frac{1+\nu}{\nu}.
\label{kappa_nu}
\end{flalign}
Inserting $\beta_{\mathrm{j}}\approx0.996$ and $\zeta\approx0.2^{\circ}$ we obtained from the analysis of VLBA data, the mass ratio is constrained as $\nu\ga0.08$. Note that if the jet inclination angle is smaller than $2\fdg2$, then $\zeta$, and consequently the lower limit on the mass ratio would become smaller.

Up to the second post-Newtonian order the merger dynamics is conservative, the constants of motion being the total energy $E$ and the angular momentum vector $\mathbf{J}=\mathbf{S}_1+\mathbf{S}_2+\mathbf{L}$. Here $\mathbf{L}$ is the orbital angular momentum, $\mathbf{S}_1$ is the spin of the dominant BH, and $\mathbf{S}_2$ is the spin of the secondary BH. The BH spins obey precessional motion \citep{Barker1975,Barker1979}:
\begin{flalign}
\mathbf{\dot{S}}_i\!=\!\Omega_i \times \mathbf{S}_i,
\end{flalign}
where $\Omega_i$ is the angular velocity of the \textit{i}-th spin $\mathbf{S}_i$. The mass ratio $\nu>0.08$ limits the ratio of the spins as $S_{1} {S_2}^{-1}<156 \chi_1 {\chi_2}^{-1}$, where $\chi_1$ and $\chi_2$ are the dimensionless spin parameters of the dominant and the secondary BHs, respectively.
In one-spin case the dominant spin $\mathbf{S}_1$ and the Newtonian angular momentum $\mathbf{L_\mathrm{N}}$ precesses around $\mathbf{J}$ with the spin-orbit period \citep{Gergely2009}:
\begin{flalign}
T_{\mathrm{SO}}\!=\!\frac{\pi Gm}{c^3} \varepsilon^{-5/2} \frac{(1+\nu)^2}{\nu}.
\label{eq:tso}
\end{flalign}
Then the dominant spin $\mathbf{S}_1$ and the orbital angular momentum $\mathbf{L}_N$ lie in the same plane, such that $\alpha=\arccos
\lbrace \mathbf{\hat{L}_\mathrm{N}} \cdot \mathbf{\hat{J}}\rbrace$, $\beta=\arccos \lbrace\mathbf{\hat{S}}_1 \cdot \mathbf{\hat{J}}\rbrace$ and $\kappa=\arccos \lbrace \mathbf{\hat{S}}_{1}\cdot \mathbf{\hat{L}_\mathrm{N}}\rbrace$ ($\kappa=\alpha+\beta$). In the typical mass ratio range of the merging SMBHs which is $1:30$ to $1:3$, a spin flip is expected at the epoch when $S_1 \approx L_\mathrm{N}$ \citep{Gergely2009}. Considering $0.08<\nu<1/3$, the ratio of the dominant spin and the Newtonian angular momentum is constrained as $0.6>S_1/L_\mathrm{N}>0.1$. In the case of PG~1302--102, this argues against the spin-flip as a viable explanation of the large difference between the inclinations of the pc- and kpc-scale jets. However, a somewhat smaller $\zeta$ value, and consequently a lower limit on the mass ratio would already allow $S_1>L_\mathrm{N}$, and therefore a spin-flip to occur (for example with $\iota=1\fdg2$, $\zeta=0\fdg1$, $\nu>0.04$).

In the inspiral stage, the approach and the eventual coalescence of the BHs is expected within the gravitational time scale of the system \citep{Gergely2009}:
\begin{flalign}
T_{\mathrm{GR}}\!=\!\frac{5Gm}{32c^{3}}\varepsilon ^{-4} \frac{(1+\nu)^2}{\nu}.
\label{eq:gr}
\end{flalign}
The mass ratio $\nu>0.08$ limits the spin-orbit period to $T_\mathrm{SO}<14\,100$~yr, and the gravitational lifetime to $T_\mathrm{GR}<7.2\times10^{6}$~yr, respectively. We summarize the binary parameters constrained by the observations of PG~1302--102 in Table~\ref{eq:binarypars}.

\begin{table}
\centering 
\caption{Binary black hole parameters derived for PG~1302--102.}
\begin{tabular}{@{}l c }
\hline
Total mass, $m^\star$ ($m_{\odot}$)  & ${\approx4\times 10^{8}}$\\
Orbital period, $T^\star$ (yr) & $4.0\pm0.2$\\
Binary separation, $r^\star$ (pc) & $\approx0.01$ \\
Post-Newtonian parameter, $\varepsilon$ & $\approx 0.002$ \\
Mass ratio, $\nu$ & $\nu>0.08$\\ 
Spin-orbit precession period, $T_{\mathrm{SO}}$ (yr) & $<14\,100$ \\
Gravitational lifetime, $T_{\mathrm{GR}}$ (yr) & $<7.2\times 10^{6}$\\
\hline 
\end{tabular}
\\
$^\star$
indicates parameters determined independently by \citet{Graham2015}.
\label{eq:binarypars}
\end{table}

\section{Summary and concluding remarks}
\label{remarks}

We analysed archival VLBA and VLA observations to reveal the pc- and kpc-scale morphology of the radio jet of PG~1302--102, a quasar possibly harbouring a tight sub-pc separation supermassive binary black hole \citep{Graham2015}. The pc-scale jet components are launched in a cone with a half-opening angle of $0\fdg20\pm0\fdg05$. However, after leaving approximately the first mas from the ``core'', they become collimated, resulting in a straight appearance of the average jet ridge line out to $\sim 4$~mas.

In our interpretation, the half-opening angle of the jet is determined by the orbital motion of the jet emitting black hole. \citet{Graham2015} found that the periodic optical variability of PG~1302--102 is consistent with varying Doppler boosting, because of the small inclination and the high speed of the pc-scale jet. They gave a set of parameters describing a jet cone, similar to what we found by analysing radio interferometric data of the pc-scale jet. \citet{Graham2015} concluded that the period most probably reveals the orbital period of the binary. Our results draw a physical picture where the orbiting black hole perturbs the jet ejection at the footpoint of the jet. This gives rise to the periodically changing jet direction and the periodical boosting of the optical flux. We constrained the binary parameters and using the VLBA data we determined a lower limit to the mass ratio of the two black holes ($\nu>0.08$).

By investigating the kpc-scale radio structure of PG~1302--102, we found that the northern part of the jet may be fitted with a helical geometric structure, suggesting a slow precession of the jet base. In this scenario, we estimated the precession period as $T_\mathrm{j} \la 10^6$~yr. However, if there is indeed a precession, the underlying physical process remains undetermined, since the large misalignment between the inclinations of the pc- and kpc-scale jets does not allow for a simple explanation of the latter within the framework of the binary model. The appearance of the kpc-scale jet in PG~1302--102, especially in the southern part, appears affected by the interaction of the jet with the dense interstellar medium. 

To reveal more about the radio structure of this peculiar source at intermediate (sub-arcsec) angular scales, interferometric imaging observation with e.g. the e-MERLIN array would be desirable. The analysis of mm-VLBI observations would be helpful to check whether an exceptionally high flux-density outburst was indicated by the MOJAVE data at the last epoch (2012.59), when the monitoring was terminated.

\section*{Acknowledgments}

EK acknowledges the support from the Campus Hungary fellowship programme, and thanks Anton Zensus for the support from the Max Planck Institute f\"ur Radioastronomie in Bonn. EK would like to thank ISSI for supporting and hosting the team on ``Unveiling multiple AGN activity in galaxy mergers''. SF and K\'EG thank for the support from the Hungarian Scientific Research Fund (OTKA, K104539). This research has made use of data from the MOJAVE database that is maintained by the MOJAVE team. The National Radio Astronomy Observatory is a facility of the National Science Foundation operated under cooperative agreement by Associated Universities, Inc.


\begin{thebibliography}{}

\bibitem[\protect\citeauthoryear{{Barker} \& {O'Connell}}{{Barker} \&
  {O'Connell}}{1975}]{Barker1975}
{Barker} B.~M.,  {O'Connell} R.~F.,  1975, Phys. Rev. D, 12, 329

\bibitem[\protect\citeauthoryear{{Barker} \& {O'Connell}}{{Barker} \&
  {O'Connell}}{1979}]{Barker1979}
{Barker} B.~M.,  {O'Connell} R.~F.,  1979, Gen. Relativ. Gravit., 11, 149

\bibitem[\protect\citeauthoryear{{Begelman}, {Blandford} \& {Rees}}{{Begelman}
  et~al.}{1980}]{Begelman1980}
{Begelman} M.~C.,  {Blandford} R.~D.,    {Rees} M.~J.,  1980, Nature, 287, 307

\bibitem[\protect\citeauthoryear{{Blandford} \& {Znajek}}{{Blandford} \&
  {Znajek}}{1977}]{BlandfordZnajek1977}
{Blandford} R.~D.,  {Znajek} R.~L.,  1977, {MNRAS}, 179, 433

\bibitem[\protect\citeauthoryear{{Caproni} \& {Abraham}}{{Caproni} \&
  {Abraham}}{2004}]{Caproni2004}
{Caproni} A.,  {Abraham} Z.,  2004, MNRAS, 349, 1218

\bibitem[\protect\citeauthoryear{{Civoni}}{{Civoni} et~al.}{2010}]{Civoni2010}
{Civoni} F. et~al.,  2010, ApJ, 717, 209 
 
\bibitem[\protect\citeauthoryear{{Condon}, {Condon}, {Gisler} \&
  {Puschell}}{{Condon} et~al.}{1982}]{Condon1982}
{Condon} J.~J.,  {Condon} M.~A.,  {Gisler} G.,    {Puschell} J.~J.,  1982, ApJ,
  252, 102

\bibitem[\protect\citeauthoryear{{Condon}, {Cotton}, {Greisen}, {Yin},
  {Perley}, {Taylor} \& {Broderick}}{{Condon} et~al.}{1998}]{Condon98}
{Condon} J.~J.,  {Cotton} W.~D.,  {Greisen} E.~W.,  {Yin} Q.~F.,  {Perley}
  R.~A.,  {Taylor} G.~B.,    {Broderick} J.~J.,  1998, AJ, 115, 1693

\bibitem[\protect\citeauthoryear{{D'Orazio}, {Haiman}, {Duffell}, {Farris} \&
  {MacFadyen}}{{D'Orazio} et~al.}{2015}]{Orazio2015}
{D'Orazio} D.~J.,  {Haiman} Z.,  {Duffell} P.,  {Farris} B.~D.,    {MacFadyen}
  A.~I.,  2015, MNRAS, 452, 2540

\bibitem[\protect\citeauthoryear{{Gergely} \& {Biermann}}{{Gergely} \&
  {Biermann}}{2009}]{Gergely2009}
{Gergely} L.~{\'A}.,  {Biermann} P.~L.,  2009, ApJ, 697, 1621

\bibitem[\protect\citeauthoryear{{Graham}, {Djorgovski}, {Stern}, {Glikman},
  {Drake}, {Mahabal}, {Donalek}, {Larson} \& {Christensen}}{{Graham}
  et~al.}{2015}]{Graham2015}
{Graham} M.~J. et al.,  2015, Nature, 518, 74

\bibitem[\protect\citeauthoryear{{Greisen}}{{Greisen}}{2003}]{Greisen03}
{Greisen} E.~W.,  2003, Information Handling in Astronomy - Historical Vistas,
  285, 109

\bibitem[\protect\citeauthoryear{{Hutchings}, {Morris}, {Gower} \&
  {Lister}}{{Hutchings} et~al.}{1994}]{Hutchings1994}
{Hutchings} J.~B.,  {Morris} S.~C.,  {Gower} A.~C.,    {Lister} M.~L.,  1994,
  PASP, 106, 642

\bibitem[\protect\citeauthoryear{{Kauffmann} \& {Haehnelt}}{{Kauffmann} \&
  {Haehnelt}}{2000}]{Kauffmann2000}
{Kauffmann} G.,  {Haehnelt} M.,  2000, MNRAS, 311, 576

\bibitem[\protect\citeauthoryear{{Kellermann}, {Sramek}, {Schmidt}, {Shaffer}
  \& {Green}}{{Kellermann} et~al.}{1989}]{Kellermann1989}
{Kellermann} K.~I.,  {Sramek} R.,  {Schmidt} M.,  {Shaffer} D.~B.,    {Green}
  R.,  1989, AJ, 98, 1195
  
\bibitem[\protect\citeauthoryear{{Komossa} \& {Zensus}}{{Komossa} \&
  {Zensus}}{2015}]{Komossa2015}
{Komossa} S.,  {Zensus} J.~A.,  2015, IAU Symp. 312, in press (arXiv:1502.05720)

\bibitem[\protect\citeauthoryear{{Kun}, {Gab{\'a}nyi}, {Karouzos}, {Britzen} \&
  {Gergely}}{{Kun} et~al.}{2014}]{Kun2014}
{Kun} E.,  {Gab{\'a}nyi} K.~{\'E}.,  {Karouzos} M.,  {Britzen} S.,    {Gergely}
  L.~{\'A}.,  2014, MNRAS, 445, 1370

\bibitem[\protect\citeauthoryear{{Lister}, {Aller}, {Aller}, {Cohen}, {Homan},
  {Kadler}, {Kellermann}, {Kovalev}, {Ros}, {Savolainen}, {Zensus} \&
  {Vermeulen}}{{Lister} et~al.}{2009}]{Lister2009}
{Lister} M.~L. et al.,  2009, AJ, 137, 3718

\bibitem[\protect\citeauthoryear{{Manchester} \& {IPTA}}{{Manchester} \&
  {IPTA}}{2013}]{Manchester2013}
{Manchester} R.~N. and  {IPTA}, 2013, Class. Quantum Grav., 30, 224010

\bibitem[\protect\citeauthoryear{{Marziani}, {Sulentic}, {Dultzin-Hacyan},
  {Calvani} \& {Moles}}{{Marziani} et~al.}{1996}]{Marziani1996}
{Marziani} P.,  {Sulentic} J.~W.,  {Dultzin-Hacyan} D.,  {Calvani} M.,
  {Moles} M.,  1996, ApJS, 104, 37

\bibitem[\protect\citeauthoryear{{Penrose}}{{Penrose}}{1969}]{Penrose1969}
{Penrose} R.,  1969, Nuovo Cimento Rivista Serie, 1, 252

\bibitem[\protect\citeauthoryear{{Planck Collaboration}}{{Planck
  Collaboration}}{2015}]{Planck2015}
{Planck Collaboration}, 2015, arXiv:1502.01589

\bibitem[\protect\citeauthoryear{{Readhead}}{{Readhead}}{1994}]{Readhead1994}
{Readhead} A.~C.~S.,  1994, ApJ, 426, 51

\bibitem[\protect\citeauthoryear{{Scheuer}}{{Scheuer}}{1995}]{Scheuer1995}
{Scheuer} P.~A.~G.,  1995, MNRAS, 277, 331

\bibitem[\protect\citeauthoryear{{Sesana}, {Vecchio} \& {Volonteri}}{{Sesana}
  et~al.}{2009}]{Sesana2009}
{Sesana} A.,  {Vecchio} A.,    {Volonteri} M.,  2009, MNRAS, 394, 2255

\bibitem[\protect\citeauthoryear{{Shepherd}, {Pearson} \& {Taylor}}{{Shepherd}
  et~al.}{1994}]{Shepherd94}
{Shepherd} M.~C.,  {Pearson} T.~J.,    {Taylor} G.~B.,  1994, BAAS, 26, 987

\bibitem[\protect\citeauthoryear{{Stocke}, {Burns} \& {Christiansen}}{{Stocke}
  et~al.}{1985}]{Stocke1985}
{Stocke} J.~T.,  {Burns} J.~O.,    {Christiansen} W.~A.,  1985, ApJ, 299, 799

\bibitem[\protect\citeauthoryear{{Stocke}, {Morris}, {Weymann} \&
  {Foltz}}{{Stocke} et~al.}{1992}]{Stocke1992}
{Stocke} J.~T.,  {Morris} S.~L.,  {Weymann} R.~J.,    {Foltz} C.~B.,  1992,
  ApJ, 396, 487

\bibitem[\protect\citeauthoryear{{Tchekhovskoy}, {Narayan} \&
  {McKinney}}{{Tchekhovskoy} et~al.}{2010}]{Tchekhovskoy2010}
{Tchekhovskoy} A.,  {Narayan} R.,    {McKinney} J.~C.,  2010, ApJ, 711, 50

\bibitem[\protect\citeauthoryear{{Wu}, {An}, {Baan}, {Hong}, {Stanghellini},
  {Frey}, {Xu}, {Liu} \& {Wang}}{{Wu} et~al.}{2013}]{Wu13}
{Wu} F. et al.,  2013, A\&A, 550, A113

\end{thebibliography}
\end{document}